%% file: conf_paper.tex
\documentclass[aps,prl,showpacs,superscriptaddress,preprint,byrevtex]{revtex4}

\usepackage{rotating}
\usepackage{longtable}
\usepackage{amsmath}
\usepackage{graphicx} % Include figure files
\usepackage{dcolumn}  % Align table columns on decimal point

\renewcommand{\arraystretch}{1.1}

\newcommand{\mev}{\,\mathrm{MeV}}
\newcommand{\mevc}{\mathrm{MeV}/c}
\newcommand{\mevm}{\mathrm{MeV}/c^2}
\newcommand{\gev}{\,\mathrm{GeV}}

\newcommand{\gevm}{\mathrm{GeV}/c^2}

\newcommand{\ee}{e^+e^-}
\newcommand{\uu}{\mu^+\mu^-}
\newcommand{\pp}{\pi^+\pi^-}
\newcommand{\ga}{\gamma}

\newcommand{\Un}{\Upsilon(nS)}
\newcommand{\Uf}{\Upsilon(5S)}
\newcommand{\Uo}{\Upsilon(1S)}
\newcommand{\Ut}{\Upsilon(2S)}
\newcommand{\Uth}{\Upsilon(3S)}

\newcommand{\mmpp}{M_{\rm miss}(\pi^+\pi^-)}
\newcommand{\mmp}{M_{\rm miss}(\pi)}

\newcommand{\et}{\eta_b(1S)}
\newcommand{\hb}{h_b(1P)}
\newcommand{\hbp}{h_b(2P)}
\newcommand{\jp}{J/\psi}
\newcommand{\ch}{\chi_{c1}}

\newcommand{\pip}{\pi^{+}}

\newcommand{\pim}{\pi^{-}}
\newcommand{\mbc}{M_{\rm bc}}

\newcommand{\de}{\Delta E}

\newcommand{\fb}{\mathrm{fb}^{-1}}
\newcommand{\br}{\mathcal{B}}

\newcommand{\etal}{\em et al.}

\newcommand{\fom}{\rm FoM}
\newcommand{\zb}{Z_b}
\newcommand{\zbo}{Z_b(10610)}
\newcommand{\zbt}{Z_b(10650)}
\newcommand{\mmppg}{M_{\rm miss}(\pi^+\pi^-\gamma)}
\newcommand{\dmmppg}{\Delta M_{\rm miss}(\pi^+\pi^-\gamma)}

\begin{document}

\title{\boldmath Observation of $\hb\to\et\gamma$}

\date{\today}

\begin{abstract}
\noindent
We report the first observation of the radiative transition
$\hb\to\et\gamma$, where the $\hb$ is produced in $\Uf\to\hb\pp$
dipion transitions. We measure the $\et$ mass to be
$(9401.0\pm1.9^{+1.4}_{-2.4})\,\mevm$ with a width of
$(12.4^{+5.5}_{-4.6}{^{+11.5}_{-3.4}})\,\mev$ and a decay branching
fraction of $\br[\hb\to\et\gamma]=(49.8\pm6.8^{+10.9}_{-5.2})\%$. The
measured $\et$ mass corresponds to a hyperfine splitting of
$(59.3\pm1.9^{+2.4}_{-1.4})\,\mevm$.  This value deviates
significantly from the current world average obtained from
measurements of $\Uth\to\et\gamma$ and $\Ut\to\et\gamma$ reactions. We
also report updated results for the $\hb$ mass
$(9899.0\pm0.4\pm1.0)\,\mevm$ and its hyperfine splitting
$(0.8\pm1.1)\,\mevm$. These measurements are performed using a
$121.4\,{\rm fb}^{-1}$ data sample collected at the peak of the $\Uf$
resonance with the Belle detector at the KEKB asymmetric-energy $\ee$
collider.
\end{abstract}

\pacs{14.40.Pq, 13.25.Gv, 12.39.Pn}

\input{author-conf2011}

\maketitle

{\renewcommand{\thefootnote}{\fnsymbol{footnote}}}
\setcounter{footnote}{0}

\section{Introduction}

Recently Belle reported the first observation of the $\hb$ and $\hbp$
states~\cite{Belle_hb}. The radiative transition to the $\et$ is
expected to be one of the dominant decay modes of the $\hb$:
$ggg/\et\gamma/gg\gamma = 57/41/2\%$~\cite{godros}. Belle's large
$\hb$ sample provide an opportunity to study the $\et$, which is the
ground state of the bottomonium system with $b\bar b$ spin and orbital
angular momentum equal to zero. The hyperfine splitting defined as
$\Delta M_{\rm HF}[\et]=M[\Uo]-M[\et]$ provides a test of spin-spin
interactions~\cite{nora}. The $\et$ was first observed by
BaBar~\cite{BaBar_etab_y3s,BaBar_etab_y2s} and confirmed by
CLEO~\cite{CLEO_etab}. Its mass is found to be higher than theoretical
predictions~\cite{pQCD,Lattice}. The tension between experimental
results and predictions strongly motivates further experimental
studies of the $\et$. We note that no experimental information is
available on the $\et$ width.

We report the first observation of the radiative transition
$\hb\to\et\gamma$ and measurements of the $\et$ mass, width and decay
branching fraction. We use a $121.4\,\fb$ data sample collected at the
peak of the $\Uf$ resonance ($\sqrt{s}\sim 10.865\gev$) with the Belle
detector~\cite{BELLE_DETECTOR} at the KEKB asymmetric-energy $\ee$
collider~\cite{KEKB}.

\section{\boldmath Method}

In the decay chain $\Uf\to\zb^+\pim$, $\zb^+\to\hb\pip$,
$\hb\to\et\gamma$ we reconstruct only the $\pim$, $\pip$ and
$\gamma$. Here $\zb^+$ denotes the $\zbo^+$ and $\zbt^+$, the two
charged bottomonium-like resonances first identified by Belle in
Ref.~\cite{Belle_zb}. In this decay the typical momenta of the $\pim$,
$\pip$ and $\gamma$ are $240\,\mevc$, $730\,\mevc$ and $500\,\mevc$,
respectively.

We define the missing mass of $X$ (where $X=\pp$ or $\pp\ga$) as 
\begin{equation}
M_{\rm miss}(X) = \sqrt{(E_{c.m.}-E_X^*)^2-p_X^{*2}},
\end{equation}
where $E_{\rm c.m.}$ is the center-of-mass (c.m.) energy and $E^*_X$
and $p^*_X$ are the energy and momentum of the $X$ system
measured in the c.m.\ frame.

\begin{figure}[htbp]
\includegraphics[width=6cm]{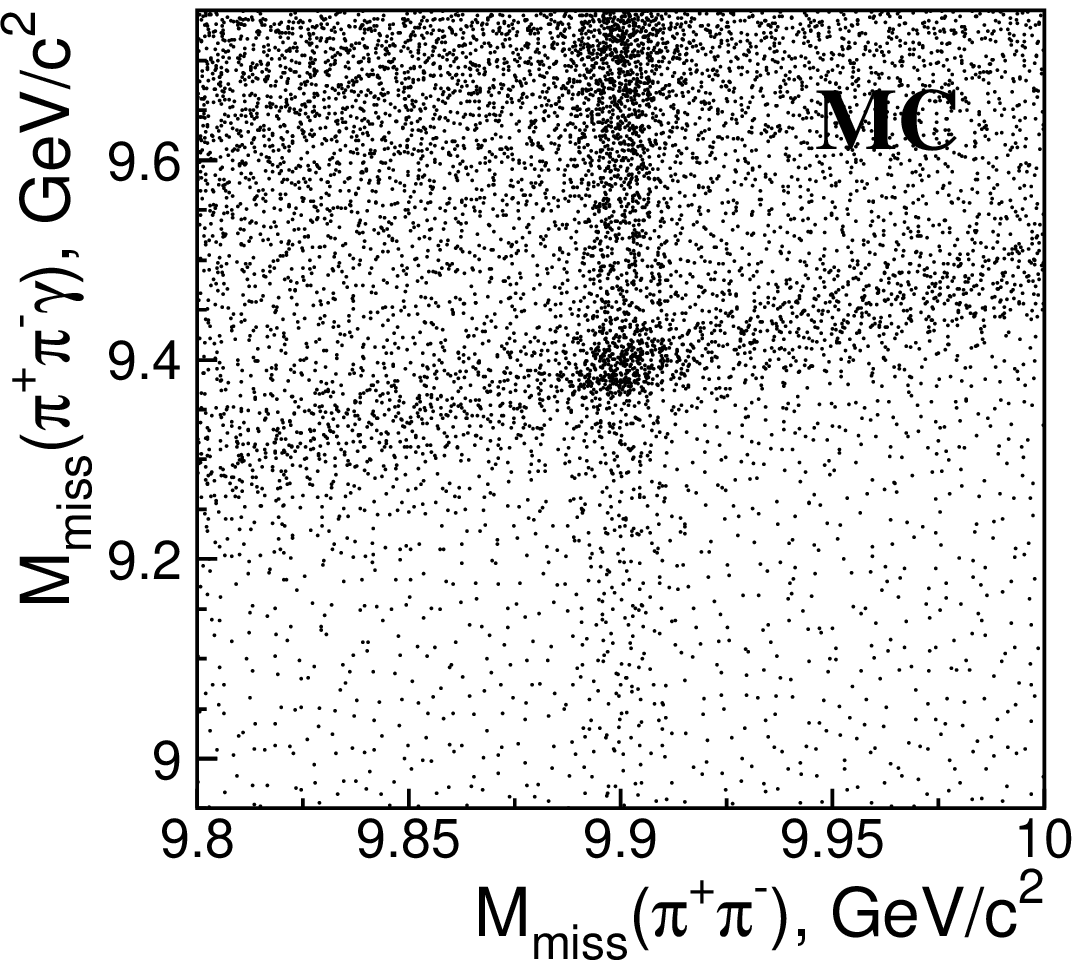}\hspace{1.5cm}
\includegraphics[width=6cm]{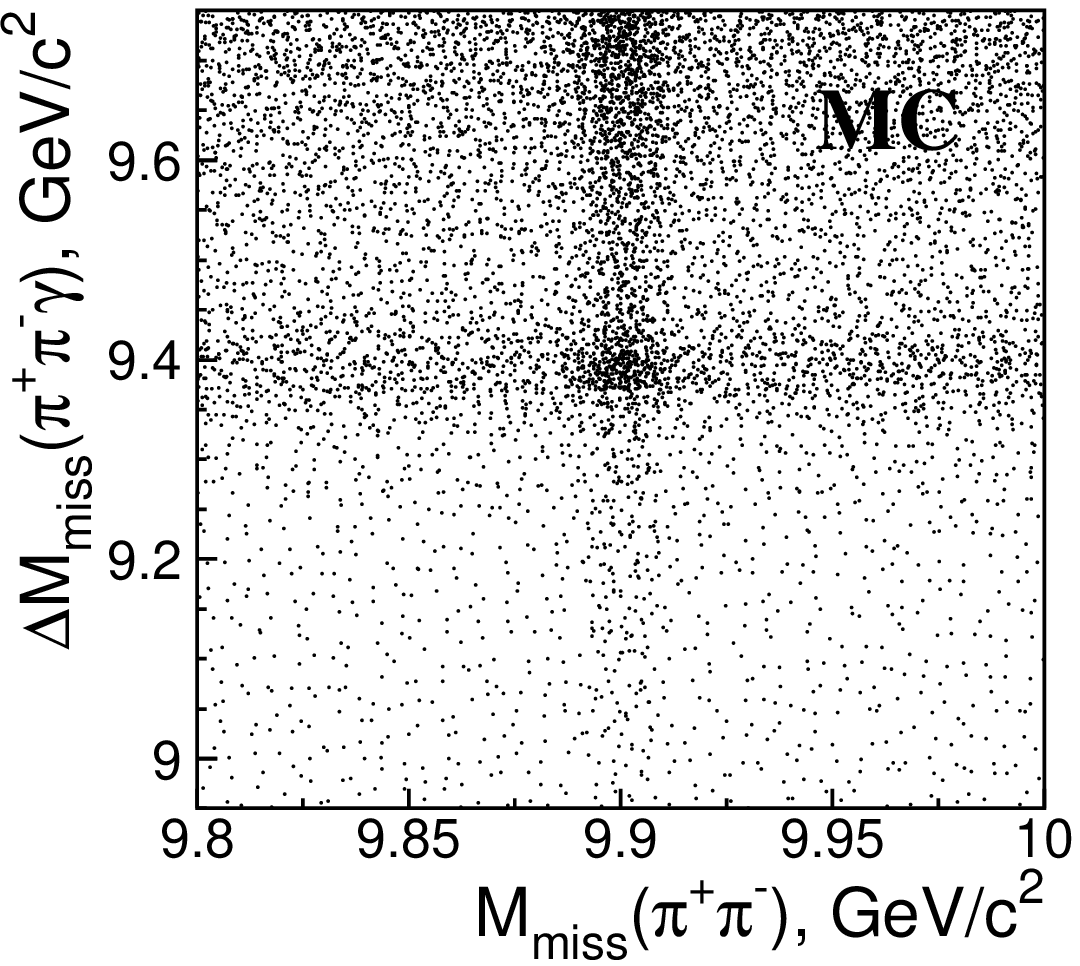}
\caption{ Results of Monte-Carlo simulation for signal $\hb\to\et\ga$
  transitions. $\mmppg$ vs. $\mmpp$ distribution (left) and $\dmmppg$
  vs. $\mmpp$ distribution (right) for all $\pp\gamma$ combinations in
  the event.}
\label{mc_mmppg_vs_mmpp}
\end{figure}

The $\pp\ga$ combinations from the signal decay chain form a cluster
in the $\mmppg$ versus $\mmpp$ plane centered at $M[\et]$ and $M[\hb]$,
respectively (see Fig.~\ref{mc_mmppg_vs_mmpp}). In this plane there is
a vertical band due to correctly reconstructed $\pp$ combinations and
misreconstructed $\ga$'s, and a slanted band due to correctly
reconstructed $\ga$'s and misreconstructed $\pp$. We introduce a new
variable $\dmmppg\equiv\mmppg-\mmpp+m[\hb]$. The advantage of the new
variable is that the band with correctly reconstructed $\ga$'s and
misreconstructed $\pp$ combinations becomes horizontal (see
Fig.~\ref{mc_mmppg_vs_mmpp}) and the variables $\dmmppg$ and $\mmpp$
are not correlated for signal events.

It may be possible to perform a two dimensional fit to the $\dmmppg$
vs. $\mmpp$ distribution. However, we follow a more intuitive
approach. We divide the $\dmmppg$ vs. $\mmpp$ plane into $10\,\mevm$
wide horizontal slices, project each slice onto the $\mmpp$ axis and
fit the $\mmpp$ distribution. We thus find the dependence of the $\hb$
yield on the $\dmmppg$ variable. 
We then search for the $\et$ signal as a peak in this distribution.

\section{\boldmath Selection}

We start with hadron event selection~\cite{hadronbj}. Continuum
$\ee\to q\bar{q}$ ($q=u,\;d,\;s,\;c$) background is suppressed by
requiring that the ratio of the second to zeroth Fox-Wolfram moments
satisfies $R_2<0.3$~\cite{Fox-Wolfram}. The $\pp$ selection
requirements are the same as in the $\hb$ and $\hbp$ observation
paper~\cite{Belle_hb}. We consider all positively identified $\pp$
pairs that originate from the vicinity of the interaction point. We
require that the $\hb$ is produced via an intermediate $\zb$,
$10.59\,\mevm<\mmp<10.67\,\mevm$~\cite{Belle_zb}.  This requirement
significantly reduces the background (by a factor of 5.2) without any
significant loss of the signal. We consider all $\gamma$ candidates,
and apply a $\pi^0$ veto, $|M(\gamma\gamma_2)-m_{\pi^0}|>13\,\mevm$
with $E_{\gamma_2}>75\,\mev$, where $\gamma_2$ is any photon candidate
in the event. The values of the cuts were optimized based on the
figure of merit $\fom=\frac{\it S}{\sqrt{\it B}}$, where $S$ is the
number of signal events in the signal Monte-Carlo (MC), $B$ is the
number of background events estimated from a small fraction (0.1\%) of
data.

\section{\boldmath Study of inclusive $\hb$ signal}

\begin{figure}[htbp]
\includegraphics[width=7cm]{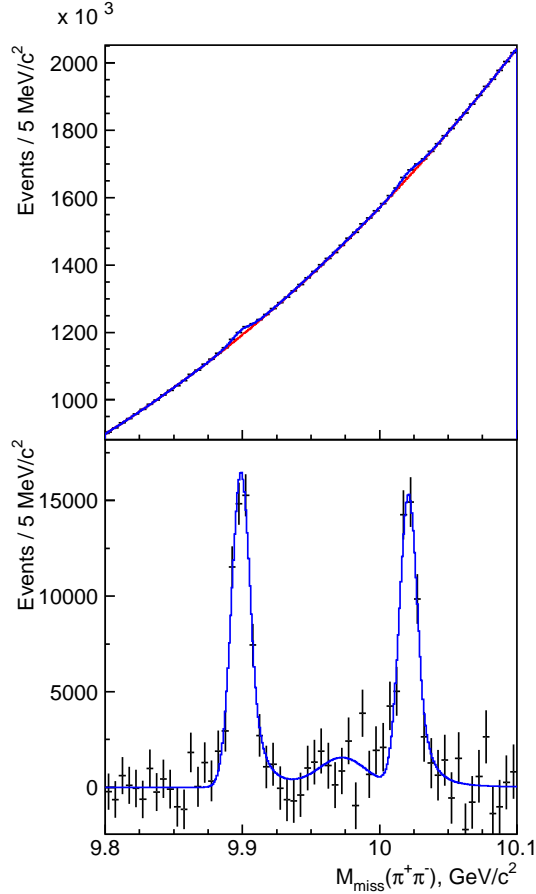}
\caption{ $\mmpp$ spectrum with the requirement of an intermediate
  $\zb$ (top) and residuals (bottom).}
\label{best_mass}
\end{figure}

The fit to the inclusive $\mmpp$ spectrum (before combining $\pp$ and
$\gamma$ candidates) with the requirement of the intermediate $\zb$ is
shown in Fig.~\ref{best_mass}.
We use the same fit procedure as in Ref.~\cite{Belle_hb}. The fit
function consists of four components: the $\hb$ signal, the $\Ut$
signal, a reflection from the $\Uth\to\Uo\pp$ decay, and combinatorial
background. The shapes of the peaking components are determined from
the analysis of the $\uu\pp$ data sample, that contains the
$\Un\to\uu$ $(n=1,2,3)$ decays. The signals are found to have tails
that account for about 8\% of the yield and are due to the initial
state radiation of soft photons. The $\hb$ and $\Ut$ intrinsic widths
are negligible compared to the detector resolution, therefore the
signals are described by a Crystal Ball function with width
$\sigma=6.5\,\mevm$ and $6.8\,\mevm$, respectively. The width
($\sigma$) of the $\hb$ signal is determined from linear interpolation
in mass from the widths of the $\Un$ peaks. The tail parameters of the
$\hb$ signal are assumed to be the same as for the $\Ut$ signal. The
$\Uth\to\Uo\pp$ reflection is described by a single Gaussian function
with a width of $\sigma=18\,\mev$. The combinatorial background is
parameterized by a third order Chebyshev polynomial. We perform binned
$\chi^2$ fit using $1\,\mevm$ bins, though for clarity we display the
data in $5\,\mevm$ bins. The results of the fit are given in
Table~\ref{tab:inclus_par}.
\begin{table}[htb]
\caption{ The yield and mass of the $\hb$ from the fit to the
  inclusive $\mmpp$ spectrum.}
\label{tab:inclus_par}
\renewcommand{\arraystretch}{1.1}
%\begin{ruledtabular}
\begin{tabular}{c|cc}
\hline
\hline
& Yield, $10^3$ & Mass, $\mevm$ \\
\hline
$\hb$          & $\;\;(61.3\pm3.1^{+2.2}_{-0.3})\times10^3\;\;\;$ & $\;(9899.0\pm0.4\pm1.0)\,\mevm$ \\
$\Uth\to\Uo\;\;$ & $(13\pm7)\times10^3$                         & $9973.0\,\mevm$ \\
$\Ut$          & $(54.8\pm3.9)\times10^3$                     & $(10021.1\pm0.5)\,\mevm$ \\
\hline
\hline
\end{tabular}
%\end{ruledtabular}
\end{table}
The confidence level of this fit is 56\%. 

To estimate systematic uncertainty on the $\hb$ parameters we vary the
Chebyshev polynomial order (+1, +2); and fit range (we reduce it to
$9.98\,\mevm$ and exclude all peaking components except for the $\hb$
signal). We also introduce a correction factor for the signal width
and allow it to float (we find $f=0.99\pm0.07$). We use a signal tail
shape not only from the $\Ut$ (the default case), but also from the
$\Uo$ and $\Uth$. A summary of the systematic uncertainties is given
in Table~\ref{tab:syst_source}.
\begin{table}[htb]
\caption{Systematic uncertainties in the $\hb$ parameters 
  from various sources. }
\label{tab:syst_source}
\renewcommand{\arraystretch}{1.0}
%\begin{ruledtabular}
\begin{tabular}{l|ccc}
\hline
\hline
& $\;\;$Polynomial$\;\;$ & Fit           & $\;\;\;$Signal$\;\;\;$ \\
& order              & $\;\;$range$\;\;$ & shape  \\
\hline
$N[\hb],\,10^3\;\;$  & $^{+1.8}_{-0}$ & $^{+1.1}_{-0}$ & $^{+0.5}_{-0.3}$ \\
$M[\hb],\,\mevm\;\;$ & $^{+0.1}_{-0}$ & $^{+0.1}_{-0}$ & $^{+0}_{-0.1}$ \\
\hline
\hline
\end{tabular}
%\end{ruledtabular}
\end{table}
For the mass measurement we introduce an additional $\pm1\,\mevm$
uncertainty due to possible local variations of background shape as
estimated in Ref.~\cite{Belle_hb} using deviations of reference
channels from the PDG values.
The new value for the $\hb$ mass corresponds to a hyperfine splitting
$\Delta M_{\rm HF}[\hb]=(0.8\pm1.1)\,\mevm$, where statistical and
systematic uncertainties are added in quadrature.

\section{\boldmath Extraction of $\et$ signal}

\begin{figure}[htbp]
\includegraphics[width=7cm]{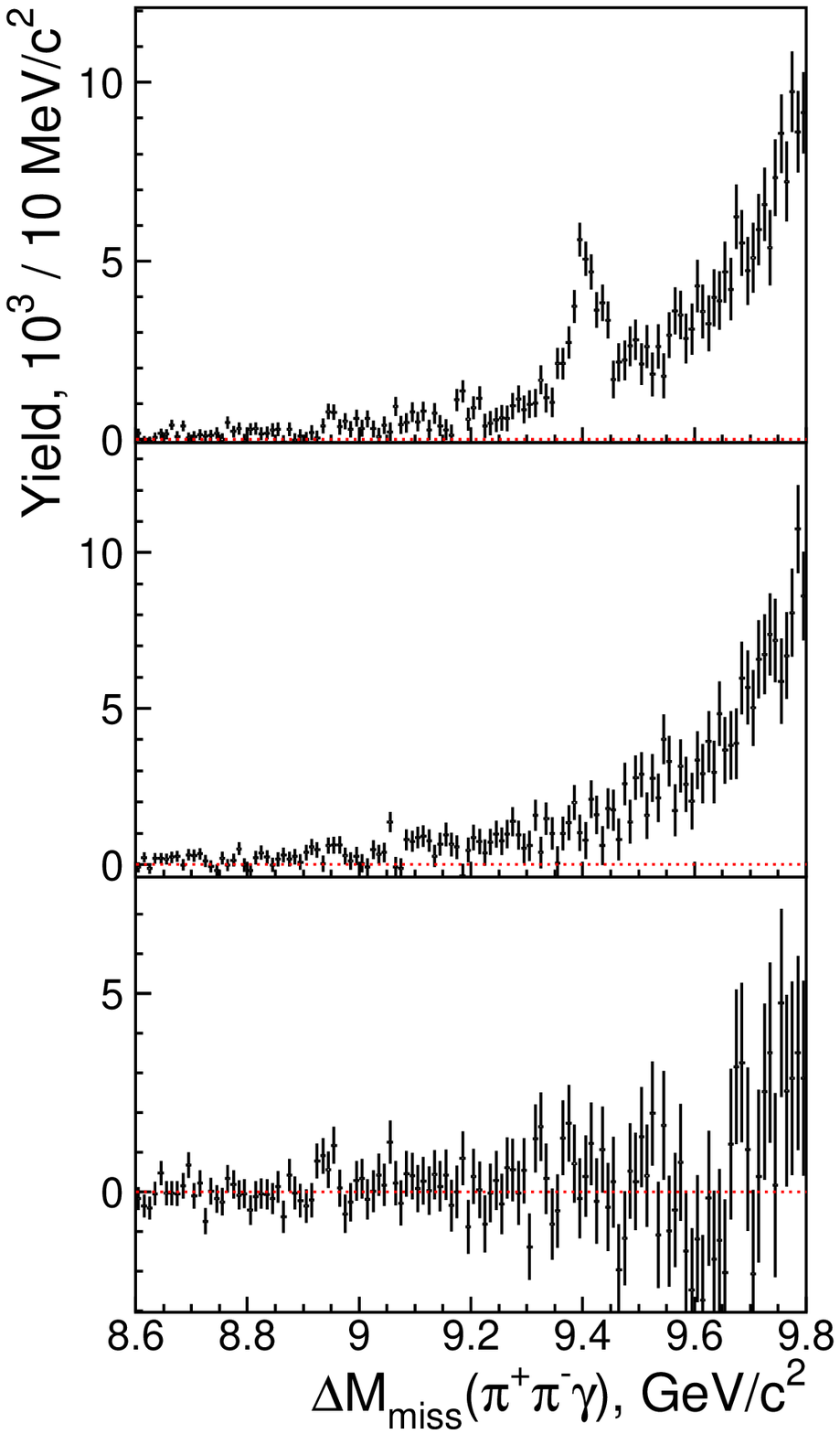}
\caption{ The results for the $\hb$ (top), $\Ut$ (middle) and
  $\Uth\to\Uo$ reflection (bottom) yields as a function of 
  $\dmmppg$. }
\label{hb_3_plots}
\end{figure}

To extract the $\et$ signal we fit the $\mmpp$ spectra in the
$\dmmppg$ bins. In the fit function we fix the masses of signals at
the values given in Table~\ref{tab:inclus_par}. We use a $10\,\mevm$
bin size in $\dmmppg$.
The results for the $\hb$, $\Ut$ and $\Uth\to\Uo$ reflection yields
as a function of the $\dmmppg$ are shown in Fig.~\ref{hb_3_plots}.
The $\hb$ distribution shows a clear peak at $9.4\,\gevm$ that we
identify as the $\et$ signal, while the other distributions do not
exhibit significant structures.

We search for peaking backgrounds in the $\dmmppg$ distribution of the
$\hb$ yield. We use a MC simulation for generic $\Uf$ decays and
consider separately $\hb\to ggg$, $\hb\to gg\gamma$ and
$e^+e^-\to\gamma_{ISR}\Uth$ processes. We do not find any sources of
peaking background.

\section{\boldmath Calibration}

\begin{figure}[htbp]
\includegraphics[width=7cm]{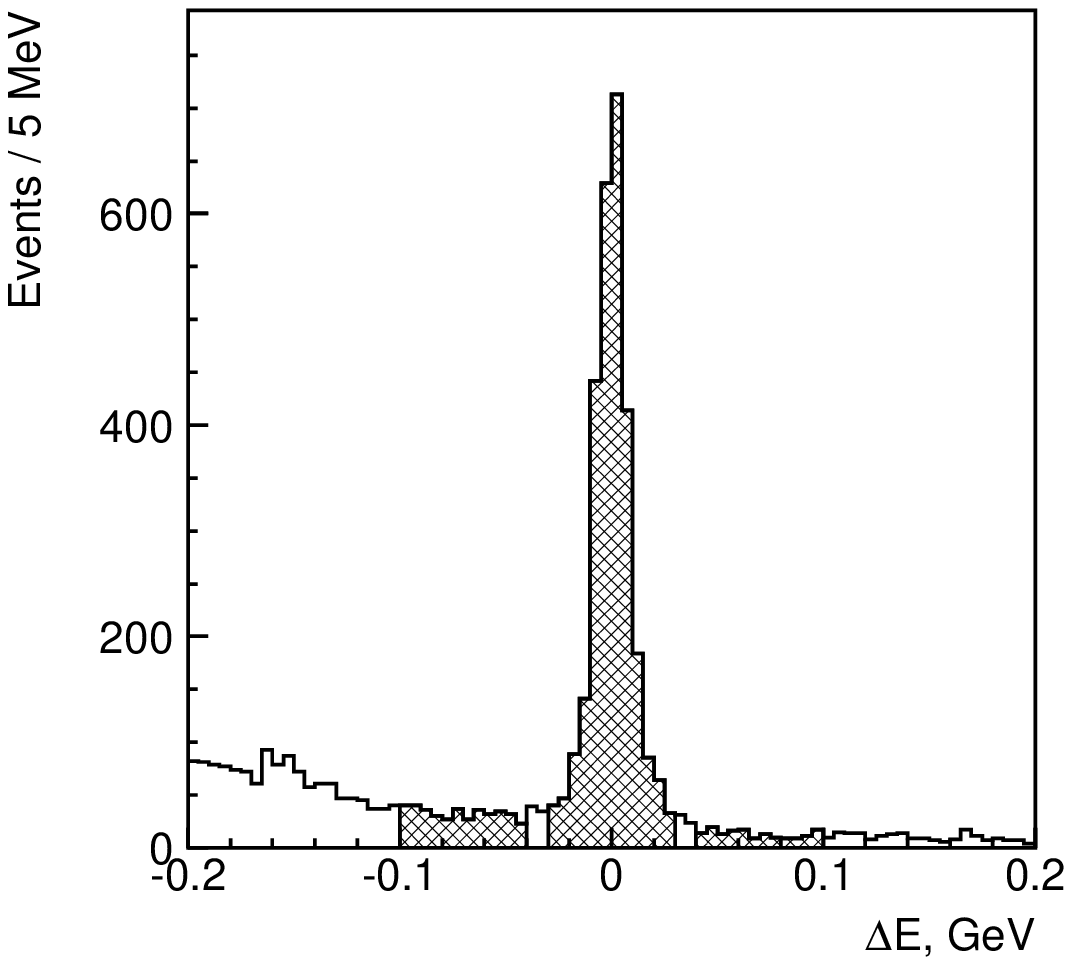}
\includegraphics[width=7cm]{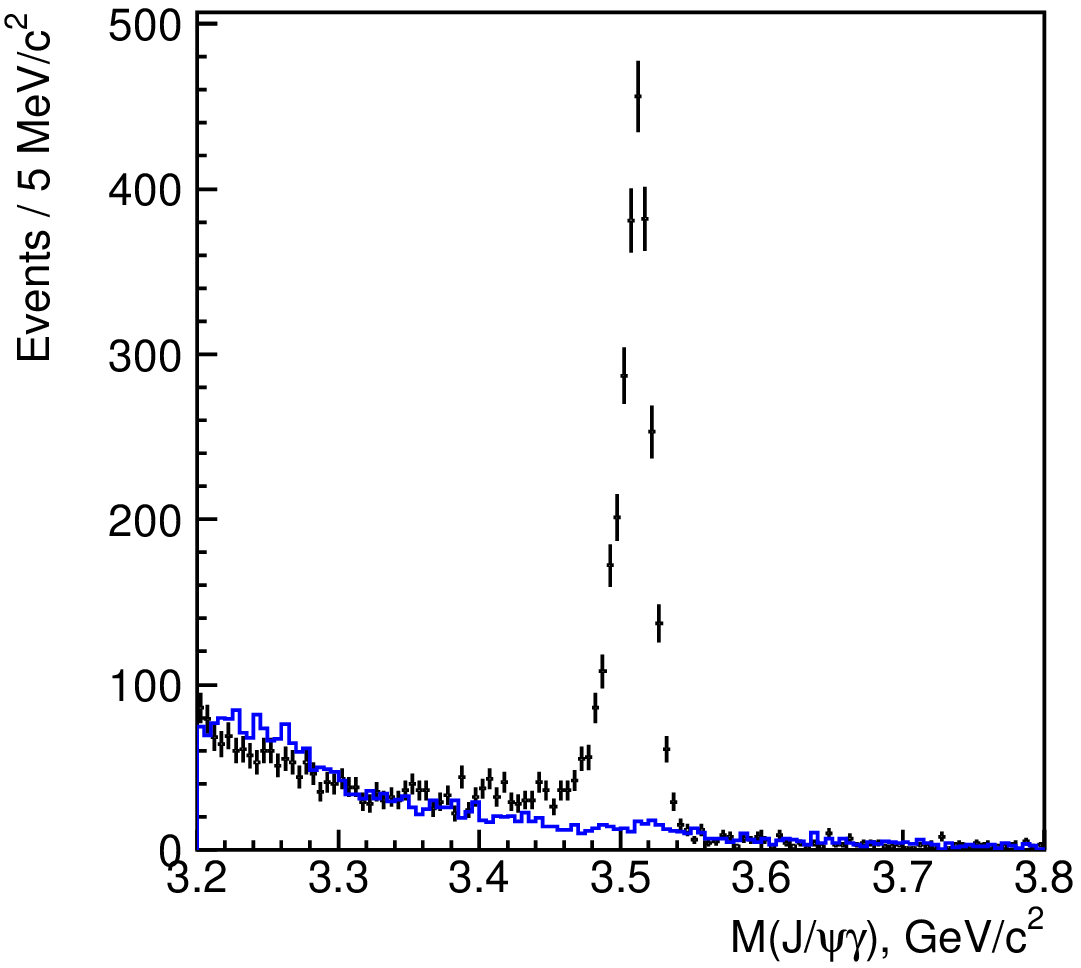}
\caption{ (Left) $\de$ distribution for the selected $B^+\to\ch K^+$
  decay candidates; signal and sidebands regions are hatched. (Right)
  $M(\jp\gamma)$ distribution for the $\de$ signal region (points with
  error bars) and for the normalized sidebands (open blue
  histogram).}
\label{de_for_chick}
\end{figure}

We use the $B^+\to\ch K^+$, $\ch\to\jp\gamma$, $\jp\to\ee/\uu$ data
sample for calibration. We require that the kaon and lepton candidates
are positively identified and originate from the vicinity of the
IP. For the $\jp\to\ee$ mode we attempt to reconstruct and recover
bremstrahlung photons. The mass window around the nominal $\jp$ mass
is $\pm30\,\mevm$ ($\pm50\,\mevm$) for the $\uu$ ($\ee$) mode.  We
perform a mass constrained fit to the $\jp$ and $\ch$ candidates.  We
require $|\Delta E|<30\,\mev$ and $\mbc>5.27\,\mevm$, which are loose
requirements. The $\Delta E$ sidebands are defined as $40<|\Delta
E|<100\,\mev$. The background is efficiently suppressed by the
requirement $\cos\theta_{\gamma}>-0.2$, where $\theta_{\gamma}$ is the
helicity angle of the $\ch$ defined as the angle between the $\gamma$
momentum and $\ch$ boost direction in the $\ch$ rest frame. The
particular value of the cut is chosen so that the average $\gamma$
energy in the c.m.\ frame is $500\,\mevm$, i.e. is equal to the
average energy of the photon in the $\hb\to\et$ transition.  The $\de$
distribution with the requirement $3.44<M(\jp\gamma)<3.54\,\gevm$ and
$M(\jp\gamma)$ distribution for the $\de$ signal region and for the
normalized sidebands (see Fig.~\ref{de_for_chick}) indicate, that the
background is small.

\begin{figure}[htbp]
\includegraphics[width=7cm]{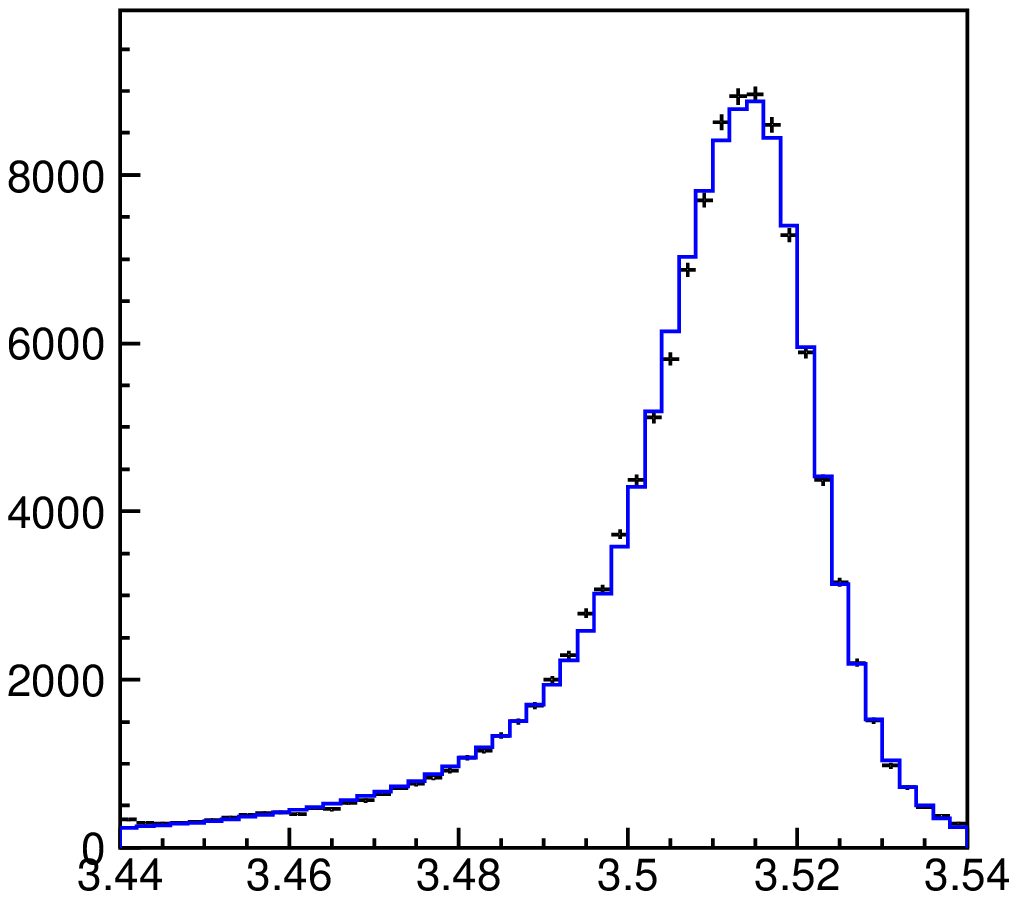}
\includegraphics[width=7cm]{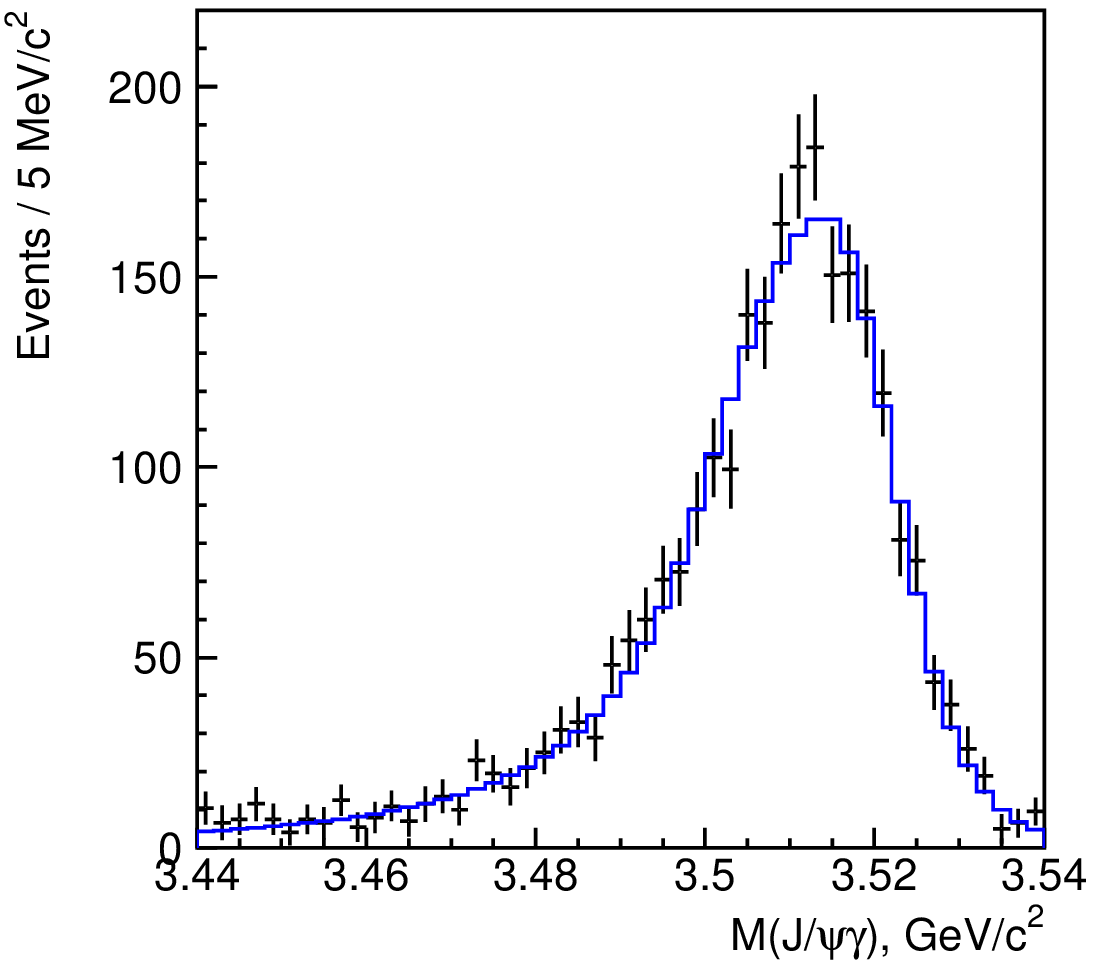}
\caption{ The $M(\jp\gamma)$ distribution in MC (left) and data $\de$
  signal region with $\de$ sidebands subtracted (right). The fit is
  described in the text.}
\label{chic_parametrization}
\end{figure}

We parameterize the $\jp\gamma$ mass distribution in MC by a Crystal
Ball function~\cite{skwarthesis} with an asymmetric core and with tails
on both sides (8 parameters in total). This parameterization describes
MC reasonably well (see Fig.~\ref{chic_parametrization}).
For the core we find $\sigma_1=11.9\,\mevm$ (left side) and
$\sigma_2=6.8\,\mevm$ (right side). 

We fit the $M(\jp\ga)$ distribution in data fixing the $\sigma_1$ and
$\sigma_2$ parameters and introducing a shift of the peak position and
width correction factor (see Fig.~\ref{chic_parametrization}). We find
for the shift: $-0.7\pm0.3^{+0.2}_{-0.4}\,\mevm$ and for the width
correction factor: $1.15\pm0.06^{+0.05}_{-0.06}$. The systematic
uncertainty is estimated (1) by varying the fit interval; (2) by using
only the left normalized $\de$ sideband for subtraction or only the
right sideband; (3) by varying the $\sigma_1/\sigma_2$ ratio in the
parameterization: we use the $\sigma_1/\sigma_2$ ratio from the fit to
the distribution in (i) data and (ii) in the $\hb\to\et\gamma$ MC.

\section{\boldmath $\et$ mass and width}

\begin{figure}[htbp]
\includegraphics[width=7cm]{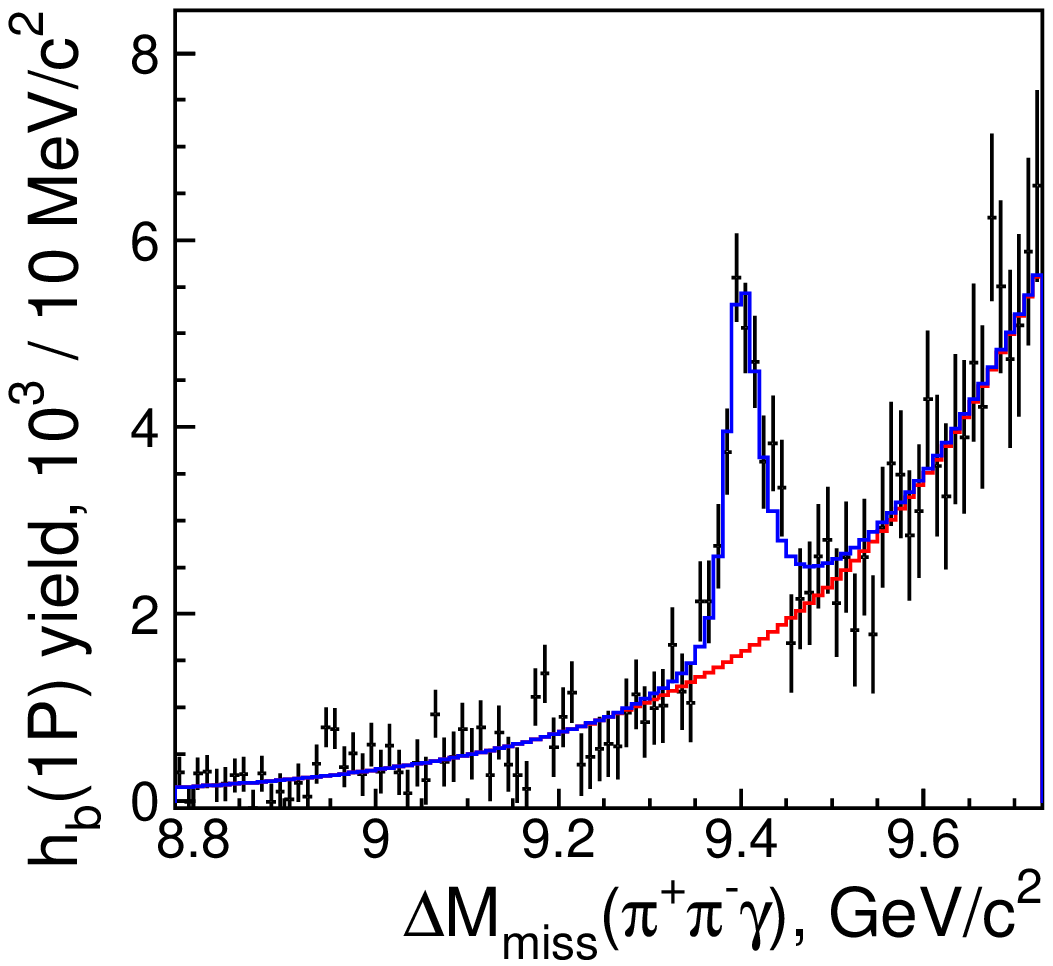}
\caption{ $\dmmppg$ distribution of the $\hb$ yield with fit result
  superimposed. }
\label{dmmppg_fit}
\end{figure}

We fit the $\dmmppg$ distribution to a sum of an $\et$ signal
component and a combinatorial background contribution (see
Fig.~\ref{dmmppg_fit}).
The signal is a non-relativistic Breit-Wigner (this parameterization
is chosen to simplify comparison with BaBar and CLEO results)
convolved with the calibrated resolution function (we use the shift
and width correction factor determined from the $\ch$ signal in
data). The combinatorial background is parameterized by an exponential
function. We verify the fit procedure using MC. The fit results are
shown in Table~\ref{tab:fit_res}.
\begin{table}[htb]
\caption{The yield, mass and width of the $\et$ signal from
  the fit to the $\dmmppg$ distribution of the $\hb$ yield. }
\label{tab:fit_res}
\renewcommand{\arraystretch}{1.1}
\begin{tabular}{l|c}
\hline
\hline
$N[\et]\;$  & $(21.9\pm2.0^{+5.6}_{-1.7})\times10^3$\\
$M[\et]\;\;$  & $\;\;(9401.0\pm1.9^{+1.4}_{-2.4})\,\mevm$\\
$\Gamma[\et]$ & $(12.4^{+5.5}_{-4.6}{^{+11.5}_{-3.4}})\,\mev$\\
\hline
\hline
\end{tabular}
\end{table}
The confidence level of the fit is 77\%. The significance of the $\et$
signal is $14\,\sigma$. 

To estimate the systematic uncertainty we vary the fit range in the
fits to the $\mmpp$ distributions (instead of the default range
$9.8-10.1\,\gevm$ we use $9.8-9.98\,\gevm$); we also vary the
polynomial order in these fits (we increase the order by one and by
two); we vary the binning of the $\dmmppg$ distribution (we scan the
starting point of bin with $1\,\mevm$ steps); we also vary the fit
range in the fits to the $\dmmppg$ distribution; we vary the
parameterization of the combinatorial background in the fits to the
$\dmmppg$ distribution (in addition to a single exponential we use the
sum of two exponentials, second and third order polynomials, we also
use the function $\exp(\sum_{k=0}^nc_kx^k)$, i.e. an exponential of a
polynomial, with $n=2,3,4$); to take into account the uncertainty in
the resolution function we vary the width correction factor by
$\pm1\,\sigma$ (we combine its statistical and systematic uncertainty
in quadrature) and we take into account the uncertainty in the mass
shift parameter; we vary the selection criteria [$R_2$ cuts: 0.25, 0.3
  (default), 0.35, 0.4; $\pi^0$ veto, cut on
  $|M(\gamma\gamma_2)-M_{\pi^0}|$: 10, 13 (default), $15\,\mevm$, cut
  on $E_{\gamma2}$: 50, 75 (default), $100\,\mev$]; we take into
account the uncertainty in the $\hb$ mass. A summary of the systematic
uncertainties is given in Table~\ref{tab:syst_et}.
\begin{table}[htb]
\caption{Systematic uncertainties in the $\et$ parameters
  from various sources. }
\label{tab:syst_et}
\renewcommand{\arraystretch}{1.0}
\begin{ruledtabular}
\begin{tabular}{l|ccc}
& $N[\et],\,10^3$ & $M[\et],\,\mevm$ & $\Gamma[\et],\,\mev$ \\
\hline
Range in $\mmpp$ fits         & $^{+0.0}_{-0.6}$ & $^{+0.0}_{-0.2}$ & $^{+0.0}_{-0.1}$ \\
Poly order in $\mmpp$ fits    & $^{+0.0}_{-0.6}$ & $^{+0.1}_{-0.1}$ & $^{+0.0}_{-0.4}$ \\
$\dmmppg$ binning             & $^{+0.2}_{-0.1}$ & $^{+0.3}_{-0.8}$ & $^{+1.0}_{-0.8}$ \\
Range in $\dmmppg$ fits       & $^{+0.9}_{-0.2}$ & $^{+0.1}_{-0.1}$ & $^{+1.4}_{-0.3}$ \\
$\dmmppg$ bg parameterization & $^{+5.5}_{-1.4}$ & $^{+0.5}_{-0.2}$ & $^{+10.9}_{-2.2}$ \\
Resolution function           & $^{+0.2}_{-0.1}$ & $^{+0.5}_{-0.5}$ & $^{+1.4}_{-1.4}$ \\
Selection requirements        & --              & $^{+0.4}_{-1.9}$ & $^{+3.0}_{-2.0}$ \\
$\hb$ mass                    & --              & $^{+1.1}_{-1.1}$ & --               \\
\hline
Total                         & $^{+5.6}_{-1.7}$ & $^{+1.4}_{-2.4}$ & $^{+11.5}_{-3.4}$ \\
\end{tabular} 
\end{ruledtabular}
\end{table}
To obtain the total systematic uncertainty we add all sources in
quadrature.

We study the shift of the $\et$ parameters in case other line-shape
parameterizations are used. We consider the KEDR
parameterization~\cite{KEDR}: $BW(m)\frac{E^3E_0^2}{EE_0+(E-E_0)^2}$,
where $BW(m)$ is the Breit-Wigner function, $E$ [$E_0$] is the
$\gamma$ energy in the $\hb$ rest frame [calculated for the pole mass
  of the $\et$]. We also consider the CLEO
parameterization~\cite{CLEO_etac}:
$BW(m)\,E^3\exp(-\frac{E^2}{8\beta^2})$, where $\beta$ is a fit
parameter. Both the KEDR and CLEO Collaborations used these
parameterizations for the $\jp\to\eta_c\gamma$ transitions.
We do not find considerable shifts in the $\et$ yield, mass or width
if these alternative parameterizations are used instead of the
non-relativistic Breit-Wigner function.

The significance of the $\et$ signal including systematic
uncertainties is $13\,\sigma$.

\section{\boldmath Measurement of $\br[\hb\to\et\gamma]$}

We measure the $\et$ [$\hb$] yield in the events that fail 
the $R_2<0.3$ or $\pi^0$ veto requirements [$R_2<0.3$ requirement]. In
the fit to the $\dmmppg$ [$\mmpp$] distribution for the rejected
events we allow the $\et$ mass and width [the $\hb$ mass] to float
within the uncertainties of the measured values given in
Table~\ref{tab:fit_res} (Table~\ref{tab:inclus_par}). We find
$N_1[\et]=(5.5\pm2.7\pm2.1)\times10^3$ and
$N_1[\hb]=(13.0\pm2.8\pm0.5)\times10^3$.

From the $N_1$ values and the yields from Tables~\ref{tab:inclus_par}
and \ref{tab:fit_res} we determine the total yields $N_0[\hb]$ and
$N_0[\et]$ without requirements on $R_2$ or a $\pi^0$ veto. We obtain
$\br=N_0[\et]/N_0[\hb]/\epsilon$, where $\epsilon$ is the
reconstruction efficiency of the radiative photon, which is found from MC to be
74.1\% with 2\% systematic uncertainty (the MC statistical uncertainty
is negligible). We find
\[
\br[\hb\to\et\gamma] = (49.8 \pm 6.8 ^{+10.9}_{-5.2})\%.
\]

\section{Conclusions}

\begin{figure}[htbp]
\includegraphics[width=8.5cm]{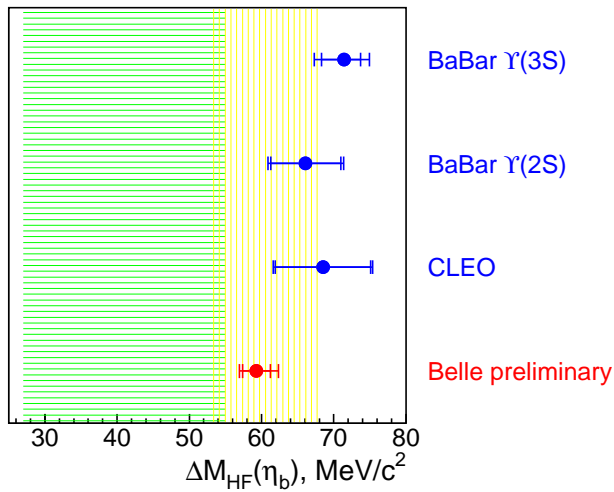}
\caption{ Hyperfine splitting measured by BaBar in $\Uth$
  data~\cite{BaBar_etab_y3s}, BaBar $\Ut$ data~\cite{BaBar_etab_y2s},
  CLEO~\cite{CLEO_etab} and present preliminary result of Belle. In
  addition, pQCD~\cite{pQCD} (horizontally hatched) and Lattice
  QCD~\cite{Lattice} (vertically hatched) predictions are shown.}
\label{exp_theory}
\end{figure}

We report the first observation of the radiative transition
$\hb\to\et\gamma$, where the $\hb$ is produced in $\Uf\to\hb\pp$
dipion transitions. We report the single most precise measurement of
the $\et$ mass, $(9401.0\pm1.9^{+1.4}_{-2.4})\,\mevm$, which
corresponds to the hyperfine splitting $\Delta M_{\rm
  HF}[\et]=(59.3\pm1.9^{+2.4}_{-1.4})\,\mevm$. This value deviates
significantly from the current world average~\cite{PDG} but decreases
tension with theoretical expectations~\cite{pQCD,Lattice} (see
Fig.~\ref{exp_theory}).
We report the first measurement of the $\et$ width
$(12.4^{+5.5}_{-4.6}{^{+11.5}_{-3.4}})\,\mev$, which is in the middle
of the range of predictions from potential models,
$4-20\,\mev$~\cite{wid_potential}. For the branching fraction we find
$\br[\hb\to\et\gamma]=(49.8\pm6.8^{+10.9}_{-5.2})\%$ in agreement with
expectations~\cite{godros}.

We also report updated results for the $\hb$ mass
$(9899.0\pm0.4\pm1.0)\,\mevm$ and hyper-fine splitting $\Delta M_{\rm
  HF}[\hb]=(0.8\pm1.1)\,\mevm$. The latter is consistent with zero, as
expected.

\end{document}

%% file: author-conf2011.tex
%%% Paper:
%%% Journal:  2011 Conference Papers
%%% March 24, 2011 - sixth draft
%%% August 29, 2011 - seventh draft (Yusa -> Niigata)
%%% Non-responding authors or those who said NO are commented out.
%%% ====================================================================
%%% Click the RELOAD button on your web browser to see the updated file.
%%% ====================================================================
%%% Use \input{author} to insert this material into your latex file.
%%%%% Force institutions to appear in alphabetical order when typeset.
\affiliation{University of Bonn, Bonn}
\affiliation{Budker Institute of Nuclear Physics SB RAS and Novosibirsk State University, Novosibirsk 630090}
\affiliation{Faculty of Mathematics and Physics, Charles University, Prague}
\affiliation{Chiba University, Chiba}
\affiliation{University of Cincinnati, Cincinnati, Ohio 45221}
\affiliation{Department of Physics, Fu Jen Catholic University, Taipei}
\affiliation{Justus-Liebig-Universit\"at Gie\ss{}en, Gie\ss{}en}
\affiliation{Gifu University, Gifu}
\affiliation{The Graduate University for Advanced Studies, Hayama}
\affiliation{Gyeongsang National University, Chinju}
\affiliation{Hanyang University, Seoul}
\affiliation{University of Hawaii, Honolulu, Hawaii 96822}
\affiliation{High Energy Accelerator Research Organization (KEK), Tsukuba}
\affiliation{Hiroshima Institute of Technology, Hiroshima}
\affiliation{University of Illinois at Urbana-Champaign, Urbana, Illinois 61801}
\affiliation{Indian Institute of Technology Guwahati, Guwahati}
\affiliation{Indian Institute of Technology Madras, Madras}
\affiliation{Indiana University, Bloomington, Indiana 47408}
\affiliation{Institute of High Energy Physics, Chinese Academy of Sciences, Beijing}
\affiliation{Institute of High Energy Physics, Vienna}
\affiliation{Institute of High Energy Physics, Protvino}
\affiliation{Institute of Mathematical Sciences, Chennai}
\affiliation{INFN - Sezione di Torino, Torino}
\affiliation{Institute for Theoretical and Experimental Physics, Moscow}
\affiliation{J. Stefan Institute, Ljubljana}
\affiliation{Kanagawa University, Yokohama}
\affiliation{Institut f\"ur Experimentelle Kernphysik, Karlsruher Institut f\"ur Technologie, Karlsruhe}
\affiliation{Korea Institute of Science and Technology Information, Daejeon}
\affiliation{Korea University, Seoul}
\affiliation{Kyoto University, Kyoto}
\affiliation{Kyungpook National University, Taegu}
\affiliation{\'Ecole Polytechnique F\'ed\'erale de Lausanne (EPFL), Lausanne}
\affiliation{Faculty of Mathematics and Physics, University of Ljubljana, Ljubljana}
\affiliation{Luther College, Decorah, Iowa 52101}
\affiliation{University of Maribor, Maribor}
\affiliation{Max-Planck-Institut f\"ur Physik, M\"unchen}
\affiliation{University of Melbourne, School of Physics, Victoria 3010}
\affiliation{Nagoya University, Nagoya}
\affiliation{Nara University of Education, Nara}
\affiliation{Nara Women's University, Nara}
\affiliation{National Central University, Chung-li}
\affiliation{National United University, Miao Li}
\affiliation{Department of Physics, National Taiwan University, Taipei}
\affiliation{H. Niewodniczanski Institute of Nuclear Physics, Krakow}
\affiliation{Nippon Dental University, Niigata}
\affiliation{Niigata University, Niigata}
\affiliation{University of Nova Gorica, Nova Gorica}
\affiliation{Osaka City University, Osaka}
\affiliation{Osaka University, Osaka}
\affiliation{Pacific Northwest National Laboratory, Richland, Washington 99352}
\affiliation{Panjab University, Chandigarh}
\affiliation{Peking University, Beijing}
\affiliation{Princeton University, Princeton, New Jersey 08544}
\affiliation{Research Center for Nuclear Physics, Osaka}
\affiliation{RIKEN BNL Research Center, Upton, New York 11973}
\affiliation{Saga University, Saga}
\affiliation{University of Science and Technology of China, Hefei}
\affiliation{Seoul National University, Seoul}
\affiliation{Shinshu University, Nagano}
\affiliation{Sungkyunkwan University, Suwon}
\affiliation{School of Physics, University of Sydney, NSW 2006}
\affiliation{Tata Institute of Fundamental Research, Mumbai}
\affiliation{Excellence Cluster Universe, Technische Universit\"at M\"unchen, Garching}
\affiliation{Toho University, Funabashi}
\affiliation{Tohoku Gakuin University, Tagajo}
\affiliation{Tohoku University, Sendai}
\affiliation{Department of Physics, University of Tokyo, Tokyo}
\affiliation{Tokyo Institute of Technology, Tokyo}
\affiliation{Tokyo Metropolitan University, Tokyo}
\affiliation{Tokyo University of Agriculture and Technology, Tokyo}
\affiliation{Toyama National College of Maritime Technology, Toyama}
\affiliation{CNP, Virginia Polytechnic Institute and State University, Blacksburg, Virginia 24061}
\affiliation{Wayne State University, Detroit, Michigan 48202}
\affiliation{Yamagata University, Yamagata}
\affiliation{Yonsei University, Seoul}
  \author{I.~Adachi}\affiliation{High Energy Accelerator Research Organization (KEK), Tsukuba} % KEK
  \author{K.~Adamczyk}\affiliation{H. Niewodniczanski Institute of Nuclear Physics, Krakow} % Krakow
  \author{H.~Aihara}\affiliation{Department of Physics, University of Tokyo, Tokyo} % Tokyo
  \author{K.~Arinstein}\affiliation{Budker Institute of Nuclear Physics SB RAS and Novosibirsk State University, Novosibirsk 630090} % BINP
  \author{Y.~Arita}\affiliation{Nagoya University, Nagoya} % Nagoya
  \author{D.~M.~Asner}\affiliation{Pacific Northwest National Laboratory, Richland, Washington 99352} % PNNL
  \author{T.~Aso}\affiliation{Toyama National College of Maritime Technology, Toyama} % Toyama
  \author{V.~Aulchenko}\affiliation{Budker Institute of Nuclear Physics SB RAS and Novosibirsk State University, Novosibirsk 630090} % BINP
  \author{T.~Aushev}\affiliation{Institute for Theoretical and Experimental Physics, Moscow} % ITEP
  \author{T.~Aziz}\affiliation{Tata Institute of Fundamental Research, Mumbai} % Tata
% ...
% BINP
\collaboration{The Belle Collaboration}